\documentclass[12pt,a4paper,twoside]{article}
\usepackage{amsmath,amsfonts,amsthm,graphicx}
\numberwithin{equation}{section} 
\newtheorem{theorem}{Theorem}[section]
\newtheorem{conjecture}[theorem]{Conjecture}
\def\beqa{\begin{eqnarray}}
\def\enqa{\end{eqnarray}}
\def\beq{\begin{equation}}
\def\enq{\end{equation}}
\def\nonum{\nonumber}
\begin{document}
\title{
 Nonlinear Integral Equations for Thermodynamics
  of the $U_{q}(\widehat{sl(r+1)})$ 
 Perk-Schultz Model
}
\author{Zengo Tsuboi 
\footnote{E-mail address: tsuboi@issp.u-tokyo.ac.jp}
and Minoru Takahashi 
\footnote{E-mail address:
mtaka@issp.u-tokyo.ac.jp}
\\
{\it Institute for Solid State Physics, University of Tokyo,} \\ 
{\it Kashiwanoha 5-1-5, Kashiwa, Chiba, 277-8581, Japan}
}
\date{}
\maketitle
\begin{abstract}
We propose a system of nonlinear integral equations (NLIE) 
which describes the thermodynamics of the 
$U_{q}(\widehat{sl(r+1)})$ Perk-Schultz model. 
These NLIE correspond to a trigonometric analogue of 
our previous result, and contain only $r$ unknown functions.
In particular, they reduce to Takahashi's NLIE for 
the $XXZ$ spin chain if $r=1$. 
We also calculate the high temperature expansion of 
the free energy. 
In particular for $r=1$ case, we have succeeded to derive 
the coefficients of order $O((\frac{J}{T})^{99})$. 
\end{abstract}
Short title: Nonlinear integral equation
{\it PACS2001:} 02.30.Rz; 02.30.Ik; 05.50.+q; 05.70.-a 
{\bf to appear in Journal of the Physical Society of Japan, Vol.74 No.3 (2005)}
\section{Introduction} 
Traditionally, thermodynamic Bethe ansatz (TBA) 
equations have been used to analyze thermodynamics of 
various kind of 
solvable lattice models \cite{Ta99}. 
However TBA equations contain, in general,  an infinite number 
of unknown functions, and are difficult to deal with. 
To overcome this difficulty, alternative 
nonlinear integral equations (NLIE), 
which contain only a finite number of unknown functions, 
have been considered. For example, for $XXZ$-model 
(or $XYZ$-model), 
Kl\"umper \cite{K93,K92}, 
Destri and de Vega \cite{DD92} 
derived a NLIE, which contains 
two unknown functions. 

Recently, the second author \cite{Ta01} derived another type of NLIE 
for the $XXZ$-model. The number of the unknown functions is 
only one, which corresponds to the rank of the 
underlying algebra $U_{q}(\widehat{sl(2)})$. 
Moreover the same NLIE was rederived \cite{TSK01} from 
the $T$-system (a 
system of functional relations among transfer matrices)
\cite{KR87,KNS94} of the quantum transfer matrix (QTM) \cite{S85}; 
from a fugacity expansion formula \cite{KW02}.  
We find \cite{ShT02} that this NLIE \cite{Ta01} 
is very useful to calculate the high temperature expansion 
of the free energy. 
In view of this situation, the first author 
derived NLIE for the $osp(1|2s)$ model \cite{T02}, 
the $sl(r+1)$ Uimin-Sutherland model \cite{T03} and the 
spin $\frac{s}{2}$ Heisenberg model \cite{T04}.  
The numbers of the unknown functions of 
these NLIE are $s,r,1$, respectively,
 which correspond to the rank of the underlying algebras.  
In particular, the NLIE in refs. \cite{T03} and \cite{T04} contain 
the rational limit of 
Takahashi's NLIE \cite{Ta01} as special cases $r=1$ or $s=1$ 
respectively. 
The high temperature expansion of the free energy 
from our NLIE \cite{T03} was applied \cite{BGOSTF03}-\cite{BGOF04} 
to spin ladder models, and good agreements were 
seen between experimental data and theoretical results. 

The Perk-Schultz model \cite{PS81,Sc83} 
is a graded vertex model
 associated with the superalgebra $U_{q}(\widehat{sl(r+1|s+1)})$. 
In this paper, we consider a special case ($s=-1$) of it: 
the $U_{q}(\widehat{sl(r+1)})$ Perk-Schultz model \cite{algebra}.  
This model is a natural multicomponent generalization of the $XXZ$ model 
 which is very important in physics. In fact the hamiltonian 
of the Perk-Schultz model (eqs. (\ref{ham0}) and (\ref{hamch})) reduces to the one 
 for the $XXZ$ model eq. (\ref{hamXXZ}) if $r=1$.  
Moreover the rational limit of the Perk-Schultz model is 
the Uimin-Sutherland model \cite{U70,S75}, which we considered 
previously \cite{T03}. 

The purpose of this paper is to derive NLIE for 
 $U_{q}(\widehat{sl(r+1)})$ Perk-Schultz model \cite{PS81,Sc83}. 
 This NLIE contains only $r$ unknown functions, 
 and corresponds to a 
 trigonometric analogue of the NLIE in ref. \cite{T03}. 
 Moreover it reduces to Takahashi's NLIE \cite{Ta01} for 
 $XXZ$-model when $r=1$. 

The shape of 
TBA equations varies with $q$ 
\cite{TS72,KSS98} 
when $q$ is a root of unity since 
the corresponding 
$T$-system (or $Y$-system),  
from which the corresponding TBA equations are derived, 
truncates with respect to a variable $m$ (cf. eq. (\ref{T-system})). 
In contrast, our new NLIE eq. (\ref{nlie4}) 
will be also valid as it stand even when $q$ is a root of unity 
since we use only the first part of the $T$-system 
(eq. (\ref{T-system}) for $m=1$). 

In ˜2, we introduce the $U_{q}(\widehat{sl(r+1)})$ 
 Perk-Schultz model \cite{PS81,Sc83} and define the 
 QTM and the $T$-system. 
In ˜3, we derive our new NLIE eq. (\ref{nlie4}) 
taking note on the periodicity of the 
QTM. We will write only the outline of the derivation 
since it is similar to the ones in refs. \cite{TSK01} and \cite{T03}. 
In ˜4, we will calculate the high temperature 
expansion of the free energy from our new NLIE. 
In particular, $r=1$ case is a detailed explanation of 
a calculation in the letter \cite{ShT02}, which has been extended to 
 the $XXZ$-model case now. 
It will be difficult to derive the same results 
from the traditional TBA equations. 
Section 5 is devoted to concluding remarks. 
\section{$T$-system and QTM Method}  
We will introduce 
the quantum transfer matrix (QTM) method 
 \cite{S85}, \cite{SI87}-\cite{SAW90}, \cite{K92,K93} 
and the $T$-system \cite{KNS94} for $U_{q}(\widehat{sl(r+1)})$ 
Perk-Schultz model. 
The QTM analyses of the Perk-Schultz model 
was done in ref. \cite{KWZ97} 
(see also, ref. \cite{JKS98,FK99}). 

The $R$-matrix of the model 
is given as 
\begin{eqnarray}
R(v)=
\sum_{a_{1}=1}^{r+1}\sum_{a_{2}=1}^{r+1}
\sum_{b_{1}=1}^{r+1}\sum_{b_{2}=1}^{r+1} 
R^{a_{1},b_{1}}_{a_{2},b_{2}}(v) 
E^{a_{1},a_{2}}\otimes E^{b_{1},b_{2}},
\end{eqnarray}
where $E^{a,b}$ is a $r+1$ by $r+1$ matrix 
whose $(i,j)$ element is given as 
$(E^{a,b})_{i,j}=\delta_{ai}\delta_{bj}$; 
$R^{a_{1},b_{1}}_{a_{2},b_{2}}(v)$ is defined as 
\begin{eqnarray}
&& R^{a,a}_{a,a}(v)=[v+1]_{q}, \\
&& R^{a,b}_{a,b}(v)=[v]_{q} \quad (a \ne b), \\
&& R^{b,a}_{a,b}(v)=q^{\mathrm{sign}(a-b)v}
\quad (a \ne b), \label{R-mat}
\end{eqnarray}
where $v \in \mathbb{C}$;
$a,b \in \{1,2,\dots,r+1\} $; $[v]_{q}=(q^{v}-q^{-v})/(q-q^{-1})$; 
$q=e^{\eta}$. 
Let $L$ be a positive integer (the number of lattice sites). 
The row-to-row transfer matrix on ${\mathbb C}_{r+1}^{\otimes L}$ 
is defined as \cite{r-index}
\begin{eqnarray}
t(v)={\mathrm tr}_{0}(R_{0L}(v)
 \cdots R_{02}(v)R_{01}(v)).
\label{rtr}
\end{eqnarray}
The hamiltonian $H=H_{0}+H_{ch}$ 
of the model has two parts. The first part $H_{0}$ 
is given by 
\begin{eqnarray}
&& \hspace{-20pt} 
H_{0}=\frac{J\sinh \eta}{\eta}\frac{d}{dv}\log t(v) |_{v=0} 
= J\sum_{j=1}^{L}\biggl\{
 \cosh \eta \sum_{a=1}^{r+1} E^{a,a}_{j}E^{a,a}_{j+1} +
\nonumber \\ && \hspace{20pt} 
 \sum_{a=1}^{r+1} 
  \sum_{
  {\scriptsize \begin{array}{c}
  b=1 \\
  a\ne b 
  \end{array}}
  }^{r+1} 
 \left( {\rm sign}(a-b) \sinh \eta 
  E^{a,a}_{j}E^{b,b}_{j+1} +
 E^{b,a}_{j}E^{a,b}_{j+1}
 \right)
\biggl\},  \label{ham0}
\end{eqnarray}
where we adopt periodic boundary condition 
$E^{a,b}_{L+1}=E^{a,b}_{1}$. 
The second part is the chemical potential term,
\begin{eqnarray}
H_{ch}=-\sum_{j=1}^{L}\sum_{a=1}^{r+1}\mu_{a}E^{a,a}_{j}.\label{hamch}
\end{eqnarray}
Let $\widetilde{R}(v)$ be \lq 90 degree rotation\rq of 
 $R(v)$:   
\begin{eqnarray}
\widetilde{R}(v)&=&
\sum_{a_{1}=1}^{r+1}\sum_{a_{2}=1}^{r+1}
\sum_{b_{1}=1}^{r+1}\sum_{b_{2}=1}^{r+1} 
\widetilde{R}^{a_{1},b_{1}}_{a_{2},b_{2}}(v) 
E^{a_{1},a_{2}}\otimes E^{b_{1},b_{2}} \nonumber \\ 
&=&
\sum_{a_{1}=1}^{r+1}\sum_{a_{2}=1}^{r+1}
\sum_{b_{1}=1}^{r+1}\sum_{b_{2}=1}^{r+1} 
R^{a_{1},b_{1}}_{a_{2},b_{2}}(v) 
^{t} \! E^{b_{1},b_{2}}\otimes E^{a_{1},a_{2}} \nonumber \\ 
&=&
\sum_{a_{1}=1}^{r+1}\sum_{a_{2}=1}^{r+1}
\sum_{b_{1}=1}^{r+1}\sum_{b_{2}=1}^{r+1} 
R^{a_{1},b_{1}}_{a_{2},b_{2}}(v) 
E^{b_{2},b_{1}}\otimes E^{a_{1},a_{2}}.
\end{eqnarray}
Namely we have $\widetilde{R}^{a_{1},b_{1}}_{a_{2},b_{2}}(v)=
R^{b_{1},a_{2}}_{b_{2},a_{1}}(v)$.
Let $N$ be a positive even integer (the Trotter number). 
We define the QTM on 
${\mathbb C}_{r+1}^{\otimes N}$ as 
\begin{eqnarray}
&& t_{\mathrm{QTM}}(v)=
{\mathrm tr}_{j} e^{- \frac{H_{j}}{T}}
R_{N,j}(u_{N}+iv)\widetilde{R}_{N-1,j}(u_{N}-iv)
\cdots  \nonumber \\
&& \hspace{45pt} 
R_{4,j}(u_{N}+iv)\widetilde{R}_{3,j}(u_{N}-iv)
R_{2,j}(u_{N}+iv)\widetilde{R}_{1,j}(u_{N}-iv),
\label{QTM}
\end{eqnarray}
where $j \in \{1,2,\dots, L \}$; 
$u_{N}=-\frac{J\sinh \eta}{\eta NT}$ 
($T$: temperature). 
The matrix elements of the QTM are given as 
\begin{eqnarray}
&& \hspace{-30pt} t_{\mathrm{QTM}}(v)=\sum_{\{\alpha_{k},\beta_{k}\}}
t_{\mathrm{QTM}}(v)
^{\{\beta_{1},\dots, \beta_{N} \}}
_{\{\alpha_{1},\dots,\alpha_{N} \}}
E^{\beta_{1}\alpha_{1}}_{1}
E^{\beta_{2}\alpha_{2}}_{2}
\cdots 
E^{\beta_{N}\alpha_{N}}_{N}, \\
&& \hspace{-46pt}
t_{\mathrm{QTM}}(v)^{\{\beta_{1},\dots, \beta_{N} \}}
_{\{\alpha_{1},\dots,\alpha_{N} \}}=
\sum_{\{\nu_{k}\}}e^{\frac{\mu_{\nu_{1}}}{T}}
\prod_{k=1}^{\frac{N}{2}}
 R^{\beta_{2k},\nu_{2k+1}}_{\alpha_{2k},\nu_{2k}}(u_{N}+iv)
 \widetilde{R}^{\beta_{2k-1},\nu_{2k}}_{\alpha_{2k-1},\nu_{2k-1}}(u_{N}-iv),
 \label{QTM2}
\end{eqnarray}
where $\nu_{N+1}=\nu_{1}$ and $\nu_{k},\alpha_{k},\beta_{k}
 \in \{1,2,\dots, r+1 \}$. 
 We can express \cite{S85} the free energy per site 
in terms of only the largest eigenvalue $\Lambda_{1}$ of 
the QTM eq. (\ref{QTM}) at $v=0$:
\begin{eqnarray}
f=
-T\lim_{N\to \infty}\log \Lambda_{1},
\end{eqnarray} 
where the Boltzmann constant is set to $1$. 

The eigenvalue formula of the row-to-row transfer matrix eq. 
(\ref{rtr}) was derived \cite{BVV82,Sc83} by the algebraic Bethe ansatz. 
On the other hand, the eigenvalue formula $T_{1}^{(1)}(v)$ 
of the QTM eq. (\ref{QTM}) is conjectured to be given by 
replacing the vacuum part (the eigenvalue on the 
pseudo-vacuum $|0>$) 
of the eigenvalue formula of the row-to-row transfer matrix 
 with that of the QTM \cite{aba}. 
 We adapt the following pseudo-vacuum: 
$|0>:=v_{r+1} \otimes v_{1} \otimes \cdots \otimes v_{r+1} \otimes v_{1} 
\in {\mathbb C}_{r+1}^{\otimes N}$, where 
 $v_{1}:=^{t}\!(1,0,\dots, 0) \in {\mathbb C}_{r+1}$ and 
$v_{r+1}:=^{t}\!(0,\dots, 0,1) \in {\mathbb C}_{r+1}$. 
Indeed $|0>$ is an eigenvector for $t_{\rm QTM}(v)$:
\begin{eqnarray} 
t_{\rm QTM}(v)|0>=\sum_{a=1}^{r+1} \psi_{a}(v)|0>,
\end{eqnarray} 
where the eigenvalue consists of the functions:
\begin{eqnarray}
&&  \hspace{-30pt} 
\psi_{a}(v)=
 e^{\frac{\mu_{a}}{T}}
 \phi_{+}(v+i \delta_{a,r+1})\phi_{-}(v-i\delta_{a,1})
 \quad 
 \mbox{for} \quad a \in \{1,2,\dots,r+1\},
 \label{vac-QTM} \\
&& \hspace{60pt} \phi_{\pm}(v)=\left(
\frac{\sin \eta (v\pm iu_{N})}{\sinh \eta }\right)^{\frac{N}{2}}.
\nonumber 
\end{eqnarray}
Then the general eigenvalue formula $T_{1}^{(1)}(v)$ 
of the QTM eq. (\ref{QTM}) will be (cf. refs. \cite{KWZ97}-\cite{FK99})
\begin{eqnarray}
T_{1}^{(1)}(v)=\sum_{d=1}^{r+1}z(d;v). 
\label{dvf-qtm}
\end{eqnarray}
Here the functions $\{z(d;v)\}$ are defined as  
\begin{eqnarray}
&& z(a;v)=\psi_{a}(v)
\frac{Q_{a-1}(v-\frac{i}{2}(a+1))Q_{a}(v-\frac{i}{2}(a-2))}
{Q_{a-1}(v-\frac{i}{2}(a-1))Q_{a}(v-\frac{i}{2}a)} 
\nonumber \\ 
&& \hspace{150pt} 
\mbox{for} \quad a \in \{1,2,\dots,r+1\}, 
\end{eqnarray}
where $Q_{a}(v)=\prod_{k=1}^{M_{a}}\sin \eta(v-v_{k}^{(a)})$; 
$M_{a}\in {\mathbb Z}_{\ge 0}$; $Q_{0}(v)=Q_{r+1}(v)=1$.  
$\{v^{(a)}_{k}\}$ is a solution of the Bethe ansatz equation 
(BAE)
\begin{eqnarray}
&& \hspace{-20pt} 
\frac{\psi_{a}(v^{(a)}_{k}+\frac{i}{2}a)}
     {\psi_{a+1}(v^{(a)}_{k}+\frac{i}{2}a)}=
-
\frac{Q_{a-1}(v^{(a)}_{k}+\frac{i}{2})Q_{a}(v^{(a)}_{k}-i)
      Q_{a+1}(v^{(a)}_{k}+\frac{i}{2})}
      {Q_{a-1}(v^{(a)}_{k}-\frac{i}{2})Q_{a}(v^{(a)}_{k}+i)
      Q_{a+1}(v^{(a)}_{k}-\frac{i}{2})}
     \label{BAE} \\
&& \hspace{40pt} \mbox{for} \quad k\in \{1,2, \dots, M_{a}\} \quad 
\mbox{and} \quad a\in \{1,2, \dots, r\}.\nonumber
\end{eqnarray}
Some remarks on eq. (\ref{dvf-qtm}) are in order. 
For $r=1$ case, eq. (\ref{dvf-qtm}) can be  
proved \cite{alge-ba} by algebraic Bethe ansatz. 
As for $r>1$ case, we have checked that 
eq. (\ref{dvf-qtm}) agrees with the 
numerical diagonalization of eq. (\ref{QTM}) for $r=2$ and 
small Trotter numbers $N$. 
In ref. \cite{KWZ97} the algebraic Bethe ansatz was executed for one particle 
state, from which  an eigenvalue formula of the QTM for the Perk-Schultz model was 
conjectured, although a different pseudo-vacuum was adopted. 

As an auxiliary function, we define 
(cf. Bazhanov-Reshetikhin formula in ref. \cite{BR90}): 
\begin{eqnarray}
T_{m}^{(a)}(v)=\sum_{\{d_{j,k}\}} \prod_{j=1}^{a}\prod_{k=1}^{m}
z(d_{j,k};v-\frac{i}{2}(a-m-2j+2k)),
\label{DVF}
\end{eqnarray}
where the summation is taken over 
$d_{j,k}\in 
\{1,2,\dots,r+1\}$ 
such that $d_{j,k} \prec d_{j+1,k}$ and $d_{j,k} \preceq d_{j,k+1}$ 
($ 1 \prec 2 \prec \cdots \prec r+1$); 
$m \in \mathbb{Z}_{\ge 1}$; 
$a \in \{1,2,\dots,r \}$. 
This function contains 
$T_{1}^{(1)}(v)$ eq. (\ref{QTM}) as a special case 
$(a,m)=(1,1)$. 
It is related to a fusion \cite{KRS81} hierarchy of the QTM. 
We can show that the poles of $T^{(a)}_{m}(v)$ 
are spurious under the BAE eq. (\ref{BAE}).  
For $a \in \{1,2,\dots,r\}$ and $m \in {\mathbb Z}_{\ge 1}$, we 
 shall normalize eq. (\ref{DVF}) as 
 $ \widetilde{T}^{(a)}_{m}(v)=
 T^{(a)}_{m}(v)/\widetilde{{\mathcal N}}^{(a)}_{m}(v)$, 
 where 
\begin{eqnarray}
\hspace{-30pt} && \widetilde{{\mathcal N}}^{(a)}_{m}(v)=
  \frac{\phi_{-}(v-\frac{a+m}{2}i)\phi_{+}(v+\frac{a+m}{2}i)}{
  \phi_{-}(v-\frac{a-m}{2}i)\phi_{+}(v+\frac{a-m}{2}i)}
  \nonumber \\ 
\hspace{-30pt}  && \hspace{20pt} \times
  \prod_{j=1}^{a}\prod_{k=1}^{m}
  \phi_{-}(v-\frac{a-m-2j+2k}{2}i)\phi_{+}(v-\frac{a-m-2j+2k}{2}i).
  \label{normal}
\end{eqnarray}
One can show that 
$\widetilde{T}^{(a)}_{m}(v)$ satisfies 
so called $T$-system \cite{KNS94}
\begin{eqnarray}
&& \hspace{-20pt}
\widetilde{T}^{(a)}_{m}(v+\frac{i}{2})
\widetilde{T}^{(a)}_{m}(v-\frac{i}{2})
=\widetilde{T}^{(a)}_{m+1}(v)\widetilde{T}^{(a)}_{m-1}(v)
+\widetilde{T}^{(a-1)}_{m}(v)\widetilde{T}^{(a+1)}_{m}(v),
\label{T-system}
\\ && \hspace{70pt} 
{\rm for} \quad a \in \{1,2,\dots,r\}
\quad {\rm and} \quad m \in {\mathbb Z}_{\ge 1}, 
\nonumber
\end{eqnarray}
where 
\begin{eqnarray}
&& \widetilde{T}^{(a)}_{0}(v)=1
\quad {\rm for} \quad a \in {\mathbb Z}_{\ge 1},\nonumber \\
&& \widetilde{T}^{(0)}_{m}(v)=
 \frac{\phi_{-}(v+\frac{m}{2}i)\phi_{+}(v-\frac{m}{2}i)}
  {\phi_{-}(v-\frac{m}{2}i)\phi_{+}(v+\frac{m}{2}i)}
\quad {\rm for} \quad m \in {\mathbb Z}_{\ge 1}, \\
&& \widetilde{T}^{(r+1)}_{m}(v)=
e^{\frac{m(\mu_{1}+\mu_{2}+\cdots +\mu_{r+1})}{T}} 
\quad {\rm for} \quad m \in {\mathbb Z}_{\ge 1} \nonumber. 
\end{eqnarray}
\section{Nonlinear Integral Equations with Only a 
Finite Number of Unknown Functions}
It is known \cite{TSK01} that 
Takahashi's NLIE \cite{Ta01}, 
which describes thermodynamics of the $XXZ$-model,  
can be derived from 
the $T$-system of the QTM. The number of unknown functions 
for this NLIE 
is only one, which corresponds to the rank of 
the underlying algebra 
$U_{q}(\widehat{sl(2)})$. 
In this section, we will derive the 
NLIE eq. (\ref{nlie4}) for $U_{q}(\widehat{sl(r+1)})$, 
which contain only $r$ unknown functions, from 
the $T$-system eq. (\ref{T-system}). 

\begin{figure}
\begin{center}
\includegraphics[width=0.95\textwidth]{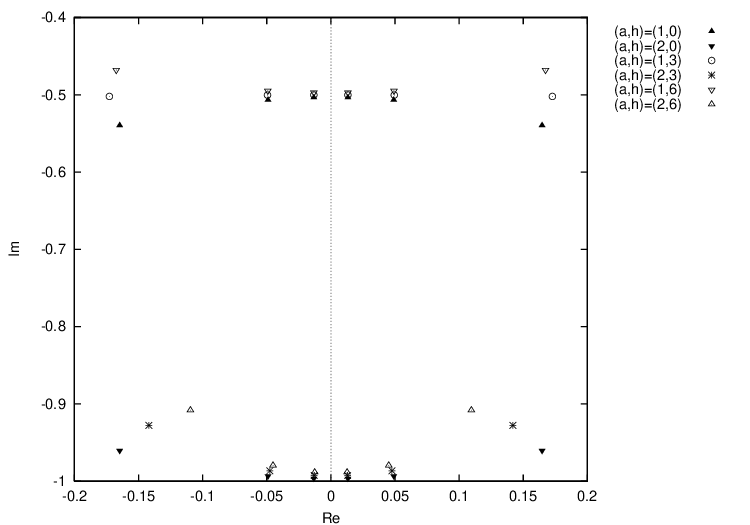}
\end{center}
\caption{Location of the roots $\{v^{(a)}_{k}\}$ ($a=1,2$) 
of the BAE for $U_{q}(\widehat{sl(3)})$ 
($\eta=2.5$, $N=12$, $u_{N}=-0.05$, 
$\mu_{1}=h,\mu_{2}=0,\mu_{3}=-h$), which will give
the largest eigenvalue of the QTM $t_{\mathrm{QTM}}(v)$ 
at $v=0$. 
Color 1 roots $\{v^{(1)}_{k}\}$ and 
color 2 roots $\{v^{(2)}_{k}\}$ 
form six one-strings. 
Only the roots in the fundamental domain 
${\rm Re} v \in [-\frac{\pi}{2\eta},\frac{\pi}{2\eta}]$ 
are exhibited. 
See also Fig. 1 in ref. \cite{T03} for $\eta=0$, $h=0$ case. }
\label{roots}
\end{figure}

\begin{figure}
\begin{center}
\includegraphics[width=0.95\textwidth]{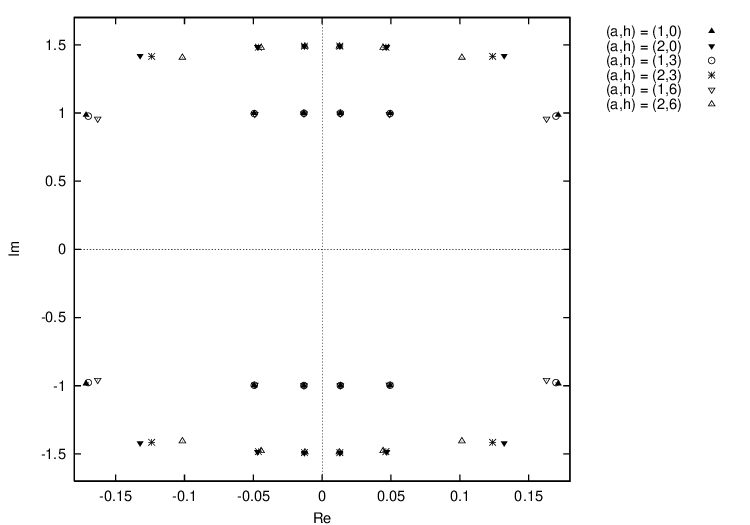}
\end{center}
\caption{Location of the zeros of $\widetilde{T}^{(a)}_{1}(v)$ 
($a=1,2$) 
corresponding to the roots in Fig. \ref{roots}. 
Only the zeros in the fundamental domain 
${\rm Re} v \in [-\frac{\pi}{2\eta},\frac{\pi}{2\eta}]$ 
are exhibited. 
See also Figs. 2 and 3 in ref. \cite{T03} for $\eta=0$, $h=0$ case.}
\label{zeros}
\end{figure}

From  a numerical analysis for finite $N,u_{N},r$, 
we conjecture that  
a one-string solution (for every color) in the sector 
$\frac{N}{2}=M_{1}=M_{2}=\cdots =M_{r}$ 
of the BAE eq. (\ref{BAE}) gives 
the largest eigenvalue of the QTM eq. (\ref{QTM}) at $v=0$. 
From now on, we will consider only this one-string solution. 
Thus $T^{(1)}_{1}(0)$ should give the 
largest eigenvalue $\Lambda_{1}$. 
We expect that 
the following conjecture is valid for 
this one-string solution 
(cf. Figs. \ref{roots} and \ref{zeros}). 
\begin{conjecture}\label{conj}
For small $u_{N}$ ($|u_{N}|\ll 1$),  
all the  zeros $\{\tilde{z}^{(a)}_{1} \}$ 
of $\widetilde{T}^{(a)}_{1}(v)$ ($a \in \{1,2,\dots,r\}$) 
are located outside of a domain (the physical strip) 
$\mathrm{Im} v \in [-\frac{1}{2},\frac{1}{2}]$. 
To be precise, they are located near the lines 
$\mathrm{ Im} v= \pm \frac{1+a}{2}$. 
\end{conjecture}
We expect that this conjecture is also valid in the 
Trotter limit $N \to \infty$. 
$\widetilde{T}^{(a)}_{m}(v)$ has poles only at 
$\pm \tilde{\beta}^{(a)}_{m,n}$: 
$\tilde{\beta}^{(a)}_{m,n}=\frac{m+a}{2}i +iu_{N}+\frac{n\pi}{\eta}$ 
($n \in \mathbb{Z}$) whose 
order is at most $N/2$. It is a periodic function with respect 
to $v$ whose 
period is $\frac{\pi }{\eta}$.
Moreover,  
\begin{eqnarray}
Q^{(a)}_{1} :=
\lim_{v\to i \eta \infty}\widetilde{T}^{(a)}_{1}(v)
=\sum_{1 \le i_{1} < i_{2} < \cdots < i_{a} \le r+1}
e^{\frac{\mu_{i_{1}}}{T}}
e^{\frac{\mu_{i_{2}}}{T}}
 \cdots
e^{\frac{\mu_{i_{a}}}{T}}
\label{Q-sl(r+1)}
\end{eqnarray}
is a finite number \cite{q-sys}. 
Thus, we must put 
\begin{eqnarray}
\widetilde{T}^{(a)}_{1}(v)=Q^{(a)}_{1} +
\sum_{n \in \mathbb{Z}}
\sum_{j=1}^{\frac{N}{2}}
\left\{
  \frac{b^{(a)}_{j}}{(v-\tilde{\beta}^{(a)}_{1,n})^{j}}
 +\frac{\overline{b}^{(a)}_{j}}{(v+\tilde{\beta}^{(a)}_{1,n})^{j}}
\right\},
\label{expan}
\end{eqnarray}
where the coefficients $b^{(a)}_{j},
\overline{b}^{(a)}_{j} \in {\mathbb C}$ are 
independent of $v$. 
Taking note on the relation 
\begin{eqnarray}
\lim_{m \to \infty}
\sum_{n=-m}^{m}\frac{1}{v-\frac{\pi n}{\eta}}
=\frac{\eta}{\tan \eta v},
\end{eqnarray}
we can derive the following NLIE for finite Trotter 
number $N$ from eqs. 
(\ref{expan}) and (\ref{T-system}) by the similar procedures 
 in refs. \cite{TSK01} and \cite{T03}. 
\begin{eqnarray}
&& \hspace{-20pt}
\widetilde{T}^{(a)}_{1}(v)=Q^{(a)}_{1} \nonumber \\ 
&& +
\oint_{C^{(a)}} \frac{{\mathrm d} y}{2\pi i} 
 \frac{\eta 
 \widetilde{T}^{(a-1)}_{1}(y+\tilde{\beta}^{(a)}_{1}-\frac{i}{2}) 
 \widetilde{T}^{(a+1)}_{1}(y+\tilde{\beta}^{(a)}_{1}-\frac{i}{2})}
 {\tan \eta (v-y-\tilde{\beta}^{(a)}_{1}) 
 \widetilde{T}^{(a)}_{1}(y+\tilde{\beta}^{(a)}_{1}-i)}
 \nonumber \\
&& +
\oint_{\overline{C}^{(a)}} \frac{{\mathrm d} y}{2\pi i} 
 \frac{
 \eta 
 \widetilde{T}^{(a-1)}_{1}(y-\tilde{\beta}^{(a)}_{1}+\frac{i}{2}) 
 \widetilde{T}^{(a+1)}_{1}(y-\tilde{\beta}^{(a)}_{1}+\frac{i}{2})}
 {\tan \eta (v-y+\tilde{\beta}^{(a)}_{1})
 \widetilde{T}^{(a)}_{1}(y-\tilde{\beta}^{(a)}_{1}+i)}
 \nonumber \\ 
 && \hspace{180pt} 
 {\rm for} \quad a \in \{1,2,\dots,r\}, \label{nlie3} 
\end{eqnarray}
where we set $\tilde{\beta}^{(a)}_{1}:=\tilde{\beta}^{(a)}_{1,0}$; 
the contour $C^{(a)}$ (resp. $\overline{C}^{(a)}$) 
is a counterclockwise closed loop around $0$ 
which satisfies the condition 
$y \ne v-\tilde{\beta}^{(a)}_{1,n}$ 
(resp. $y \ne v+\tilde{\beta}^{(a)}_{1,n}$) and 
does not surround 
$\tilde{z}^{(a)}_{1}-\tilde{\beta}^{(a)}_{1,n}+i$, 
$-2\tilde{\beta}^{(a)}_{1}+\frac{\pi n}{\eta}$, $\frac{\pi k}{\eta}$ 
(resp. 
$\tilde{z}^{(a)}_{1}+\tilde{\beta}^{(a)}_{1,n}-i$, 
$2 \tilde{\beta}^{(a)}_{1}+\frac{\pi n}{\eta}$, $\frac{\pi k}{\eta}$) 
($n \in \mathbb{Z}$, $k \in \mathbb{Z}-\{0\}$). 

Next we will take the Trotter limit $N \to \infty$. 
We put $\mathcal{T}^{(a)}_{1}(v):=
\lim_{N \to \infty} \widetilde{T}^{(a)}_{1}(v)$. 
In particular, we have 
\begin{eqnarray} 
\mathcal{T}^{(0)}_{1}(v)=
\lim_{N \to \infty} \widetilde{T}^{(0)}_{1}(v)
=\exp \left(\frac{2J (\sinh \eta)^{2} }
{T(\cosh \eta -\cos (2\eta v))}\right).
\end{eqnarray}
Then, we arrive at a system of NLIE, which
contains only a {\em finite} number of unknown 
functions $\{{\mathcal T}^{(a)}_{1}(v) \}_{1\le a \le r}$: 
\begin{eqnarray}
{\mathcal T}^{(a)}_{1}(v)=Q^{(a)}_{1} 
&+&
\oint_{C^{(a)}} \frac{{\mathrm d} y}{2\pi i} 
 \frac{\eta 
 \mathcal{T}^{(a-1)}_{1}(y+\frac{i a}{2}) 
 \mathcal{T}^{(a+1)}_{1}(y+\frac{i a}{2})}
 {\tan \eta (v-y-\frac{i(a+1)}{2})
 \mathcal{T}^{(a)}_{1}(y+\frac{i(a-1)}{2})}
 \nonumber \\
&+&
\oint_{\overline{C}^{(a)}} \frac{{\mathrm d} y}{2\pi i} 
 \frac{\eta 
 \mathcal{T}^{(a-1)}_{1}(y-\frac{i a}{2}) 
 \mathcal{T}^{(a+1)}_{1}(y-\frac{i a}{2})}
 {\tan \eta (v-y+\frac{i(a+1)}{2})
 \mathcal{T}^{(a)}_{1}(y-\frac{i(a-1)}{2})}
 \nonumber \\ 
 && \hspace{120pt} 
 {\rm for} \quad a \in \{1,2,\dots,r\},
 \label{nlie4}
\end{eqnarray}
where $\mathcal{T}^{(r+1)}_{1}(v)=Q^{(r+1)}_{1}=
e^{\frac{\mu_{1}+\cdots +\mu_{r+1}}{T}}$; 
the contour $C^{(a)}$ (resp. $\overline{C}^{(a)}$) 
is a counterclockwise closed loop around $0$ 
which satisfies the condition 
$y \ne v-\beta^{(a)}_{1,n}$ 
(resp. $y \ne v+\beta^{(a)}_{1,n}$) and 
does not surround 
$z^{(a)}_{1}-\beta^{(a)}_{1,n}+i$, 
$-(a+1)i
+\frac{\pi n}{\eta}$, $\frac{\pi k}{\eta}$ 
(resp. 
$z^{(a)}_{1}+\beta^{(a)}_{1,n}-i$, 
$(a+1)i
+\frac{\pi n}{\eta}$, $\frac{\pi k}{\eta}$); 
$\beta^{(a)}_{1,n}=\lim_{N\to \infty}\tilde{\beta}^{(a)}_{1,n}
=\frac{a+1}{2}i+\frac{\pi n}{\eta}$;  
$z^{(a)}_{1}=\lim_{N\to \infty }{\tilde z}^{(a)}_{1}$ 
($n \in \mathbb{Z}$, $k \in \mathbb{Z}-\{0\}$). 
Note that eq. (\ref{nlie4}) reduces 
to Takahashi's NLIE \cite{Ta01} if $r=1$. 
Although we assumed that 
$q=e^{\eta}$ is not a root of unity, 
we think that eq. (\ref{nlie4}) will be also valid as it stand 
even when $q$ is a root of unity. 
One can calculate the free energy per site $f$ 
by using eq. (\ref{nlie4}) 
and the relation
\begin{eqnarray}
f=J \cosh \eta -T\log \mathcal{T}^{(1)}_{1}(0). 
\label{free-en}
\end{eqnarray}
\section{High Temperature Expansion}
In this section, we will calculate the high temperature 
expansion of the free energy eq. (\ref{free-en}) for 
$U_{q}(\widehat{sl(r+1)})$ 
by our new NLIE eq. (\ref{nlie4}).  
It is known that 
the high temperature expansion of the free energy 
for the $XXX$-model 
was calculated \cite{ShT02} by Takahashi's NLIE  
up to order 100. 

First of all, we assume the following expansion for large $T/|J|$: 
\begin{eqnarray}
\mathcal{T}^{(a)}_{1}(v)=
 \exp \left(\sum_{n=0}^{\infty}b_{n}^{(a)}(v)
 (\frac{J}{T})^{n} \right), 
 \quad a \in \{1,2,\dots, r\},
 \label{t-expan}
\end{eqnarray}
where $b_{0}^{(a)}(v)=\log Q^{(a)}_{1}$. 
 Substituting eq. (\ref{t-expan}) into eq. (\ref{nlie4}), we 
can obtain the coefficients $\{b_{n}^{(a)}(v)\}$. 
We treat $r=1$ case separately. \\ 
$\bullet $ $r=1$ case:\\ 
In this case, the Hamiltonian eqs. (\ref{ham0}) and (\ref{hamch}) becomes XXZ chain 
in magnetic field:
\beqa
&&H_0+H_{ch}= 2J\sum_{j=1}^L \Bigl\{ S_j^xS_{j+1}^x+S_j^yS_{j+1}^y+
\Delta (S_j^zS_{j+1}^z+\frac{1}{4}) \Bigr\} \nonum\\
&& \hspace{20pt} -L(\mu_1+\mu_2)-2
(\mu_1-\mu_2)\sum_{j=1}^L S_j^z, \label{hamXXZ}
\enqa
where $\Delta=\cosh \eta$. 
The NLIE has especially simple form. 
Equation (\ref{nlie4}) is 
\beqa
&& \mathcal{T}^{(1)}_{1} (v)= Q^{(1)}_{1} +Q^{(2)}_{1} \oint_C 
\Bigl(\frac{\mathcal{T}^{(0)}_{1}(y+i/2)}{\tan{\eta}(v-y-i)}
+\frac{\mathcal{T}^{(0)}_{1}(y-i/2)}{\tan{\eta}(v-y+i)}
\Bigr){\eta\over \mathcal{T}^{(1)}_{1}(y)}{{\rm d}y\over 2\pi i }\nonumber 
\\ \label{NLIEsl2}
\enqa
Contour $C$ is a closed loop counterclockwise around $0$. 
\noindent
To remove trigonometric functions, we use transformation
\beq
v=F(X), \quad y=F(Y): 
\quad F(X)=\frac{1}{\eta}\tan^{-1}\Bigl({iX\sinh\eta\over \sqrt{1+(\cosh\eta)^{2} X^{2}}}\Bigr).
\enq
By this transformation, points $(0,\pm i, \infty)$ on $v$ plane move to $(0, \infty, \pm i)$ 
on $X$ plane, respectively.
 Then eq. (\ref{NLIEsl2}) becomes  
\beqa
&&\mathcal{T}^{(1)}_{1} (F(X))=Q^{(1)}_{1} +Q^{(2)}_{1} (1+X^2) \oint_C A(X,Y)
\frac{1}{\mathcal{T}^{(1)}_{1} (F(Y))}\frac{dY}{2\pi i},\nonum\\
&&A(X,Y)=\sum_{\epsilon=\pm 1}\frac{\frac{Y(1-X^2Y^2+2\Delta^2X^2(1+Y^2))}{1+Y^2}
+\epsilon\frac{\Delta(1+(1-2\Delta^2)X^2Y^2)}{\sqrt{1+\Delta^2 Y^2}}}
{(1-X^2Y^2)^2-4\Delta^2 X^2 Y^2(1+X^2)(1+Y^2)}\nonum\\
&&~~~~~~~~~~\times\exp\bigl\{\frac{J}{T}(1+Y^2)(\Delta+\epsilon\frac{\sqrt{1+\Delta^2Y^2}}{Y})\bigr\},
\enqa
where we assume that $\mathcal{T}^{(1)}_{1} (F(Y))$ is an even function of $Y$. 
New integration kernel $A$ is expanded by power series of $J/T$. Coefficient of 
$l$-th term is
\beqa
&&
\frac{1}{ l!}\frac{(1+Y^2)^l}
{(1-X^2Y^2)^2-4\Delta^2 X^2 Y^2(1+X^2)(1+Y^2)}\nonum\\
&&\Bigl(\frac{Y(1-X^2Y^2+2\Delta^2X^2(1+Y^2))}{1+Y^2}
((\Delta+\frac{\sqrt{1+Y^2\Delta^2}}{Y})^l+(\Delta-\frac{\sqrt{1+Y^2\Delta^2}}{Y})^l)\nonum\\
&&+\frac{\Delta(1+(1-2\Delta^2)X^2Y^2)}{\sqrt{1+\Delta^2 Y^2}}
((\Delta+\frac{\sqrt{1+Y^2\Delta^2}}{Y})^l-(\Delta-\frac{\sqrt{1+Y^2\Delta^2}}{Y})^l)
\Bigr).\label{exA}
\enqa
Analytic part of function $A$ at $Y=0$ is not relevant in this integral. Then we can put
\beqa
&&A=({\rm analytic~~ part~~at~~}Y=0)+ \sum_{l=1}^\infty{J^l\over l!T^l}A_l,
\enqa
We get $A_l$ from eq. (\ref{exA}) as polynomials of $X^2, \Delta^2$ and $1/Y$;
{\beqa
&&A_1=\frac{\Delta }{Y},\quad
A_2=\frac{1 + 2\,{\Delta }^2 + 2{\Delta }^2X^2}{Y},\nonum\\
&&A_3=\frac{\Delta }{Y^3} + \frac{6\Delta + 
            4{\Delta }^3  + (3\Delta  + 8{\Delta }^3)X^2 + 4 {\Delta }^3X^4}{Y},\quad 
A_4=\frac{1 + 4{\Delta }^2 + 2 {\Delta }^2X^2}{Y^3}\nonum\\
&& + \frac{3 + 24{\Delta }^2 + 8{\Delta }^4 + (1+ 24{\Delta }^2 + 24{\Delta }^4)X^2  
+ 8({\Delta }^2 + 24{\Delta }^4) X^4+ 8{\Delta }^4X^6}{Y},\nonum\\
&&....
\enqa
}
We can determine $b^{(1)}_{l}(F(X))$ successively from the integral equation. 
These are polynomials of $X^2, \Delta$ and ${Q^{(2)}_{1}}/{{Q^{(1)}_{1}}^2}$
($=1/(2 \cosh \frac{h}{T})^{2}$, $h=(\mu_{1}-\mu_{2})/2$ ). 
\begin{eqnarray}
&& \hspace{-30pt} b^{(1)}_{1}(0)=
\frac{2\,\Delta\,Q^{(2)}_{1}}{{Q^{(1)}_{1}}^2}, 
\label{hte-1} \\
&& \hspace{-30pt} b^{(1)}_{2}(0)=
\frac{\left( 1 + 2\,\Delta^2 \right) \,
Q^{(2)}_{1}}{{Q^{(1)}_{1}}^2} - 
\frac{6\,\Delta^2\,{Q^{(2)}_{1}}^2}{{Q^{(1)}_{1}}^4}, 
\label{hte-2} \\
&& \hspace{-30pt} b^{(1)}_{3}(0)=
\frac{\left( 2\,\Delta + \frac{4\,\Delta^3}{3} \right) \,
 Q^{(2)}_{1}}{{Q^{(1)}_{1}}^2} + \frac{\left( -6\,\Delta - 12\,\Delta^3 \right)
           \,{Q^{(2)}_{1}}^2}{{Q^{(1)}_{1}}^4} + 
              \frac{80\,\Delta^3\,{Q^{(2)}_{1}}^3}{3\,{Q^{(1)}_{1}}^6},   
              \label{hte-3} \\
&& \hspace{-30pt} b^{(1)}_{4}(0)=
\frac{\left( \frac{1}{4} + 2\,\Delta^2 + \frac{2\,\Delta^4}{3} \right) \,
Q^{(2)}_{1}}{{Q^{(1)}_{1}}^2} 
 + \frac{\left( -\frac{3}{2}  - \
\frac{56\,\Delta^2}{3} - 14\,\Delta^4 \right) 
\,{Q^{(2)}_{1}}^2}{{Q^{(1)}_{1}}^4} 
\nonumber \\ &&
\hspace{50pt} + \frac{\left(40\,\Delta^2 + 80\,\Delta^4 \right) \,
{Q^{(2)}_{1}}^3}{{Q^{(1)}_{1}}^6} - 
\frac{140\,\Delta^4\,{Q^{(2)}_{1}}^4}{{Q^{(1)}_{1}}^8}.
 \label{hte-4}
\end{eqnarray}
In Table \ref{coet}, we give $ j!b^{(1)}_{j}(0)$ up to $j=10$. 
\begin{table} 
\caption{ $ j!\cdot b^{(1)}_{j}(0)$ for $j=1,...,10$.  
We put $M={Q^{(2)}_{1}}/{{Q^{(1)}_{1}}^2}$ 
($=1/(2 \cosh \frac{h}{T})^{2}$, $h=(\mu_{1}-\mu_{2})/2$ ).}
{\tiny 
\begin{eqnarray}
1 &&2M\Delta ,\quad\nonum\\
2 &&2M + 4M{\Delta }^2 - 12M^2{\Delta }^2, \nonum\\
3 &&
12\,M\,\Delta  - 36\,M^2\,\Delta  + 8\,M\,{\Delta }^3 - 
  72\,M^2\,{\Delta }^3 + 160\,M^3\,{\Delta }^3, \nonum\\
4 &&
6\,M - 36\,M^2 + 48\,M\,{\Delta }^2 - 448\,M^2\,{\Delta }^2 + 
  960\,M^3\,{\Delta }^2 + 16\,M\,{\Delta }^4 - 336\,M^2\,{\Delta }^4\nonum\\
&&
 + 
  1920\,M^3\,{\Delta }^4 - 3360\,M^4\,{\Delta }^4, \nonum\\
5 &&
60\,M\,\Delta  - 960\,M^2\,\Delta  + 2400\,M^3\,\Delta  + 
  160\,M\,{\Delta }^3 - 3600\,M^2\,{\Delta }^3 + 20000\,M^3\,{\Delta }^3 - 
  33600\,M^4\,{\Delta }^3 \nonum\\
&&
+ 32\,M\,{\Delta }^5 - 1440\,M^2\,{\Delta }^5 + 
  16000\,M^3\,{\Delta }^5 - 67200\,M^4\,{\Delta }^5 + 96768\,M^5\,{\Delta }^5, \nonum\\
6 &&
20\,M - 600\,M^2 + 2400\,M^3 + 360\,M\,{\Delta }^2 - 
  13320\,M^2\,{\Delta }^2 + 85200\,M^3\,{\Delta }^2 - 
  151200\,M^4\,{\Delta }^2\nonum\\
&&
 + 480\,M\,{\Delta }^4 - 23808\,M^2\,{\Delta }^4 + 
  259200\,M^3\,{\Delta }^4 - 1048320\,M^4\,{\Delta }^4 + 
  1451520\,M^5\,{\Delta }^4 \nonum\\
&&
+ 64\,M\,{\Delta }^6 - 5952\,M^2\,{\Delta }^6 + 
  115200\,M^3\,{\Delta }^6 - 873600\,M^4\,{\Delta }^6 + 
  2903040\,M^5\,{\Delta }^6\nonum\\
&&
 - 3548160\,M^6\,{\Delta }^6, \nonum\\
7 &&
280\,M\,\Delta  - 18760\,M^2\,\Delta  + 168000\,M^3\,\Delta  - 
  352800\,M^4\,\Delta  + 1680\,M\,{\Delta }^3\nonum\\
&&
 - 135408\,M^2\,{\Delta }^3 + 
  1689856\,M^3\,{\Delta }^3 - 7197120\,M^4\,{\Delta }^3 + 
  10160640\,M^5\,{\Delta }^3 \nonum\\
&&
+ 1344\,M\,{\Delta }^5 - 
  141120\,M^2\,{\Delta }^5 + 2696960\,M^3\,{\Delta }^5 - 
  19756800\,M^4\,{\Delta }^5 \nonum\\
&&
+ 63221760\,M^5\,{\Delta }^5 - 
  74511360\,M^6\,{\Delta }^5 + 128\,M\,{\Delta }^7 - 
  24192\,M^2\,{\Delta }^7 \nonum\\
&&
+ 770560\,M^3\,{\Delta }^7 - 
  9408000\,M^4\,{\Delta }^7 + 54190080\,M^5\,{\Delta }^7 - 
  149022720\,M^6\,{\Delta }^7 \nonum\\
&&
+ 158146560\,M^7\,{\Delta }^7, \nonum\\
8 &&
70\,M - 8820\,M^2 + 117600\,M^3 - 352800\,M^4 + 2240\,M\,{\Delta }^2 - 
  322560\,M^2\,{\Delta }^2 \nonum\\
&&
+ 5388096\,M^3\,{\Delta }^2 - 
  26530560\,M^4\,{\Delta }^2 + 40642560\,M^5\,{\Delta }^2 + 
  6720\,M\,{\Delta }^4 \nonum\\
&&
- 1145088\,M^2\,{\Delta }^4 + 
  25079040\,M^3\,{\Delta }^4 - 193277952\,M^4\,{\Delta }^4 + 
  629959680\,M^5\,{\Delta }^4 \nonum\\
&&
- 745113600\,M^6\,{\Delta }^4 + 
  3584\,M\,{\Delta }^6 - 780288\,M^2\,{\Delta }^6 + 
  24729600\,M^3\,{\Delta }^6 \nonum\\
&&
- 292626432\,M^4\,{\Delta }^6 + 
  1625702400\,M^5\,{\Delta }^6 - 4314562560\,M^6\,{\Delta }^6 \nonum\\
&&
+ 
  4428103680\,M^7\,{\Delta }^6 + 256\,M\,{\Delta }^8 - 
  97536\,M^2\,{\Delta }^8 + 4945920\,M^3\,{\Delta }^8 \nonum\\
&&
- 
  91445760\,M^4\,{\Delta }^8 + 812851200\,M^5\,{\Delta }^8 - 
  3775242240\,M^6\,{\Delta }^8 + 8856207360\,M^7\,{\Delta }^8 \nonum\\
&&
- 
  8302694400\,M^8\,{\Delta }^8, \nonum\\
9 &&
1260\,M\,\Delta  - 337680\,M^2\,\Delta  + 8023680\,M^3\,\Delta  - 
  50803200\,M^4\,\Delta  + 91445760\,M^5\,\Delta \nonum\\
&&
 + 13440\,M\,{\Delta }^3 - 
  4072320\,M^2\,{\Delta }^3 + 116998560\,M^3\,{\Delta }^3 - 
  1030458240\,M^4\,{\Delta }^3 \nonum\\
&&
+ 3617187840\,M^5\,{\Delta }^3 - 
  4470681600\,M^6\,{\Delta }^3 + 24192\,M\,{\Delta }^5 - 
  8570880\,M^2\,{\Delta }^5 \nonum\\
&&
+ 311921280\,M^3\,{\Delta }^5 - 
  3884186880\,M^4\,{\Delta }^5 + 21973690368\,M^5\,{\Delta }^5 - 
  58502062080\,M^6\,{\Delta }^5 \nonum\\
&&
+ 59779399680\,M^7\,{\Delta }^5 + 
  9216\,M\,{\Delta }^7 - 4112640\,M^2\,{\Delta }^7 + 
  209088000\,M^3\,{\Delta }^7 \nonum\\
&&
- 3759436800\,M^4\,{\Delta }^7 + 
  32286449664\,M^5\,{\Delta }^7 - 144850083840\,M^6\,{\Delta }^7 + 
  328786698240\,M^7\,{\Delta }^7\nonum\\
&&
 - 298896998400\,M^8\,{\Delta }^7 + 
  512\,M\,{\Delta }^9 - 391680\,M^2\,{\Delta }^9 + 
  30976000\,M^3\,{\Delta }^9 - 835430400\,M^4\,{\Delta }^9 \nonum\\
&&
+ 
  10762149888\,M^5\,{\Delta }^9 - 75107450880\,M^6\,{\Delta }^9 + 
  292254842880\,M^7\,{\Delta }^9 \nonum\\
&&
- 597793996800\,M^8\,{\Delta }^9 + 
  501851750400\,M^9\,{\Delta }^9, \nonum\\
10 &&
252\,M - 128520\,M^2 + 4445280\,M^3 - 38102400\,M^4 + 91445760\,M^5 + 
  12600\,M\,{\Delta }^2\nonum\\
&&
 - 7072800\,M^2\,{\Delta }^2 + 
  282582720\,M^3\,{\Delta }^2 - 3083330880\,M^4\,{\Delta }^2 + 
  12395980800\,M^5\,{\Delta }^2\nonum\\
&&
 - 16765056000\,M^6\,{\Delta }^2 + 
  67200\,M\,{\Delta }^4 - 42297600\,M^2\,{\Delta }^4 + 
  1999721280\,M^3\,{\Delta }^4 \nonum\\
&&
- 28282381440\,M^4\,{\Delta }^4 + 
  171577405440\,M^5\,{\Delta }^4 - 475808256000\,M^6\,{\Delta }^4 + 
  498161664000\,M^7\,{\Delta }^4\nonum\\
&&
 + 80640\,M\,{\Delta }^6 - 
  58803840\,M^2\,{\Delta }^6 + 3444222720\,M^3\,{\Delta }^6 - 
  65241899520\,M^4\,{\Delta }^6 \nonum\\
&&
+ 570726051840\,M^5\,{\Delta }^6 - 
  2568513024000\,M^6\,{\Delta }^6 + 5803583385600\,M^7\,{\Delta }^6 \nonum\\
&&
- 
  5230697472000\,M^8\,{\Delta }^6 + 23040\,M\,{\Delta }^8 - 
  20930560\,M^2\,{\Delta }^8 + 1671936000\,M^3\,{\Delta }^8 \nonum\\
&&
- 
  44003635200\,M^4\,{\Delta }^8 + 548674560000\,M^5\,{\Delta }^8 - 
  3702575923200\,M^6\,{\Delta }^8\nonum\\
&&
 + 13948526592000\,M^7\,{\Delta }^8 - 
  27675648000000\,M^8\,{\Delta }^8 + 22583328768000\,M^9\,{\Delta }^8 \nonum\\
&&
+ 
  1024\,M\,{\Delta }^{10} - 1569792\,M^2\,{\Delta }^{10}+ 
  191078400\,M^3\,{\Delta }^{10} - 7333939200\,M^4\,{\Delta }^{10} + 
  131681894400\,M^5\,{\Delta }^{10} \nonum\\
&&
- 1295901573120\,M^6\,{\Delta }^{10} + 
  7439214182400\,M^7\,{\Delta }^{10} - 24908083200000\,M^8\,{\Delta }^{10}\nonum\\
&&
 + 
  45166657536000\,M^9\,{\Delta }^{10} - 
  34326659727360\,M^{10}\,{\Delta }^{10} \nonum
\end{eqnarray}
}
\label{coet}
\end{table} 

\begin{figure}
\begin{center}
\includegraphics[width=0.75\textwidth]
{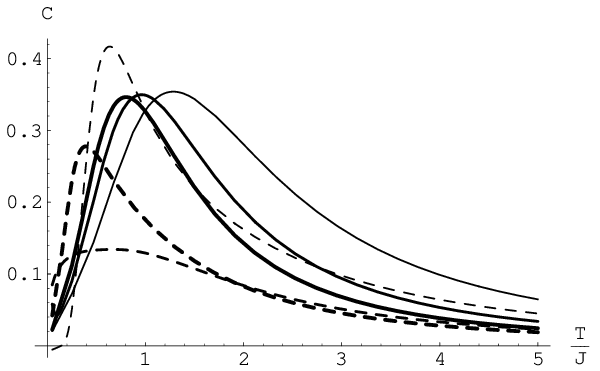}
\end{center}
\caption{Temperature dependence of the specific heat $C$ 
for $U_{q}(\widehat{sl(2)})$ ($\mu_{1}=-\mu_{2}=0$) calculated 
from Pade approximation 
(numerator: a polynomial of degree 49, 
denominator: a polynomial of degree 50)
for $\Delta=$2 (thin), 1 (medium), 0.5 (thick), 
-0.5 (dashed thick), -1 (dashed medium), -2 (dashed thin). 
We have calculated high-temperature expansion of free energy up to 99-th order. 
  The case for $\Delta=\pm 1$ coincides with ref. \cite{ShT02}.}
\label{specific-h=2}
\end{figure}

\begin{figure}
\begin{center}
\includegraphics[width=0.75\textwidth]
{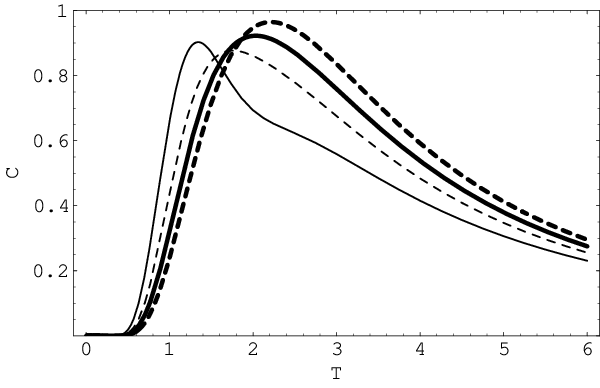}
\end{center}
\caption{Temperature dependence of the high temperature 
expansion of the specific heat $C$ 
for $U_{q}(\widehat{sl(4)})$ ($J=0.48$, $\mu_{1}=6.75$, 
$\mu_{2}=\mu_{3}=\mu_{4}=0$): 
 each line denotes $C$ for 
$\eta=\frac{3}{2}$ (thin),$1$ (dashed thin), $0$ (thick), 
$\frac{\pi i}{3}$ (dashed thick) respectively.  
 The case for $\eta=0$  
 was calculated in ref. \cite{BGOSTF03} in relation 
 with the $su(4)$ ladder model.}
\label{specific-sl4}
\end{figure}

There are a lot of works on high temperature expansions for the XXX model;
 only a few works exist for the XXZ model.
Destri and de Vega 's \cite{DV95} result contains some misprint. 
Rojas et al \cite{RST02} also calculated the coefficients 
up to order $O((\frac{J}{T})^{4})$. Unfortunately, their formula 
(eqs. (46)-(48) in ref. \cite{RST02}) 
contains a misprint of a sign. 
We note that eqs. 
(\ref{hte-1})-(\ref{hte-3}) agree with  eq. (46) in ref. \cite{RST02}. 
In addition, eq. (\ref{hte-4}) also agrees with  eq. (46) in ref. 
\cite{RST02} {\em if} 
we replace the term $-\frac{t^4}{32}\beta^3$ in eq. (48)  
with $+\frac{t^4}{32}\beta^3$. 
Moreover, these coefficients eqs. (\ref{hte-1})-(\ref{hte-4}) reduce to 
known results (see for example, refs. \cite{ShT02} and \cite{BEU00}) 
in the rational limit $\Delta \to 1$. 
We have plotted specific heat by using our high temperature 
expansion formula of order 99 (see Fig. \ref{specific-h=2}). \\
$\bullet $ $r \ge 2$ case: \\ 
This case is more involved than $r=1$ case since 
we have to take account of 
the poles in the coefficients $\{b_{n}^{(a)}(v)\}$. Thus 
we need a further non-trivial assumption 
\begin{eqnarray}
b_{n}^{(a)}(v)=\left(\frac{2 (\sinh \eta)^{2}}
{\cosh (a+1)\eta -\cos 2\eta v}\right)^n c^{(a)}_{n}(v),
\end{eqnarray}
where $c_{n}^{(a)}(v)$ is a polynomial 
with respect to some trigonometric functions. 
We will present $b^{(1)}_{n}(0)$ for $U_{q}(\widehat{sl(r+1)})$ up 
to order $n=3$. 
\begin{eqnarray}
&& b^{(1)}_{1}(0)=
\frac{2\,\Delta\,Q^{(2)}_{1}}{{Q^{(1)}_{1}}^2}, 
\label{hte-hr1} \\
&& b^{(1)}_{2}(0)= 
\frac{\left( 1 + 2\,\Delta^2 \right) \,Q^{(2)}_{1}}{{Q^{(1)}_{1}}^2} - 
  \frac{6\,\Delta^2\,{Q^{(2)}_{1}}^2}{{Q^{(1)}_{1}}^4} + 
  \frac{\left( -1 + 4\,\Delta^2 \right) \,Q^{(3)}_{1}}{{Q^{(1)}_{1}}^3},
  \label{hte-hr2}
\end{eqnarray}
\begin{eqnarray}
&& b^{(1)}_{3}(0)=
\frac{\left( 2\,\Delta + \frac{4\,\Delta^3}{3} \right) 
\,Q^{(2)}_{1}}{{Q^{(1)}_{1}}^2} + 
  \frac{\left( -6\,\Delta - 12\,\Delta^3 \right) 
  \,{Q^{(2)}_{1}}^2}{{Q^{(1)}_{1}}^4} + 
  \frac{80\,\Delta^3\,{Q^{(2)}_{1}}^3}{3\,{Q^{(1)}_{1}}^6} 
  \nonumber \\ && \hspace{10pt} + 
  \frac{8\,\Delta^3\,Q^{(3)}_{1}}{{Q^{(1)}_{1}}^3} + 
  \frac{\left( 8\,\Delta - 32\,\Delta^3 \right) 
  \,Q^{(2)}_{1}\,Q^{(3)}_{1}}{{Q^{(1)}_{1}}^5} + 
  \frac{\left( -4\,\Delta + 8\,\Delta^3 \right) 
  \,Q^{(4)}_{1}}{{Q^{(1)}_{1}}^4}. \label{hte-hr3}
\end{eqnarray}
We note the following relation: 
$b_{n}^{(1)}(0)=c_{n}^{(1)}(0)$. 
We can obtain coefficients for $r=2$ if we formally 
set $Q^{(4)}_{1}=0$ in eq. (\ref{hte-hr3}). 
Note that eqs. (\ref{hte-hr1}), (\ref{hte-hr2}), (\ref{hte-hr3}) 
reduce to eqs. 
(\ref{hte-1}), (\ref{hte-2}), (\ref{hte-3}) 
if $Q^{(3)}_{1}=Q^{(4)}_{1}=0$. 
Note also that eqs. (\ref{hte-hr1})-(\ref{hte-hr3}) reduce to 
 eq. (5.6) in ref. \cite{T03} in the rational limit $\Delta \to 1$. 
We have plotted the specific heat by using our high temperature 
expansion formula eqs. (\ref{hte-hr1})-(\ref{hte-hr3}) of order 3 
 (see Fig. \ref{specific-sl4}). 
\section{Concluding Remarks}
We have derived NLIE with only $r$ unknown functions 
for the $U_{q}(\widehat{sl(r+1)})$ Perk-Schultz model. 
In contrast with traditional TBA equations, 
our new NLIE will be also valid even when $q$ is a root of unity. 
We also calculated the high temperature expansion 
of the free energy from our new NLIE. 

It will be an important problem to apply our result  
to physical problem such as the thermodynamics of 
the spin ladder models as in ref. \cite{BGOSTF03}. 
In Ref. \cite{YRFC04}, they treated a spin ladder model whose 
leg part has isotropic interaction  
 while rung part has anisotropic interaction. 
In contrast, our new NLIE has a potential 
applicability to spin ladder models 
whose leg parts have anisotropic interaction. 

We note that our results in this paper reduce to the ones in 
ref. \cite{T03} in the rational limit $q=e^{\eta} \to 1$. 
\section*{Acknowledgments}
\noindent
This work is partially supported by 
Grant-in-Aid for Scientific Research 
from JSPS (no. 16914018). 
  

\begin{thebibliography}{80}
\bibitem{Ta99} 
M. Takahashi: Thermodynamics of One-Dimensional Solvable 
 models, (Cambridge University Press, Cambridge, 1999).

\bibitem{K93} 
  A. Kl\"umper: Z. Phys. {\bf B91} (1993) 507-519.

\bibitem{K92} 
     A. Kl\"umper: 
        Ann. Physik {\bf 1} (1992) 540-553.

\bibitem{DD92}
C. Destri and H. J. de Vega: 
Phys. Rev. Lett. {\bf 69} (1992) 2313-2317.

\bibitem{Ta01} 
M. Takahashi: in Physics and Combinatorics, 
eds. A. N. Kirillov and N. Liskova, (2001) 299--304 
(World Scientific, Singapore); 
cond-mat/0010486.

\bibitem{TSK01}
M. Takahashi, M. Shiroishi and A. Kl\"umper: 
J. Phys. A: Math. Gen. {\bf 34} (2001) L187--L194; 
cond-mat/0102027.

\bibitem{KR87} 
        A. N. Kirillov and N. Yu. Reshetikhin:   
        J. Phys. A: Math. Gen. {\bf 20} (1987) 1565--1585.

\bibitem{KNS94}
A. Kuniba, T. Nakanishi and J. Suzuki:
Int. J. Mod. Phys. {\bf A9} (1994) 5215-5266; 
hep-th/9309137. 

\bibitem{S85} 
        M. Suzuki: Phys. Rev. {\bf B31} (1985) 2957--2965.

\bibitem{KW02}
 G. Kato and M. Wadati:
 J. Math. Phys. {\bf 43} (2002) 5060--5078; 
 cond-mat/0212325.

\bibitem{ShT02}
M. Shiroishi and M. Takahashi:  
Phys. Rev. Lett. 89 (2002) 117201; cond-mat/0205180.

\bibitem{T02}
 Z. Tsuboi:  
        Phys. Lett. {\bf B544} (2002) 222--230; 
        math-ph/0209024.                           

\bibitem{T03}
 Z. Tsuboi: 
 J. Phys. A: Math. Gen. {\bf 36} (2003) 1493--1507; 
 cond-mat/0212280.

\bibitem{T04}
Z. Tsuboi:  
J. Phys. A: Math. Gen. 
{\bf 37} (2004) 1747--1758; cond-mat/0308333. 

\bibitem{BGOSTF03} 
M. T. Batchelor, X. W. Guan, N. Oelkers, 
K. Sakai, Z. Tsuboi, A. Foerster: 
Phys. Rev. Lett. {\bf 91} (2003) 217202; 
cond-mat/0309244.

\bibitem{YRFC04}
Zu-Jian Ying, I. Roditi, A. Foerster, 
 B. Chen: 
Euro. Phys. J. {\bf B41} (2004) 67-74;  
cond-mat/0403520.
 
\bibitem{YRZ04}
Zu-Jian Ying, I. Roditi, Huan-Qiang Zhou: 
cond-mat/0405274. 

\bibitem{BGO04}
M. T. Batchelor, X. W. Guan, N. Oelkers: 
Phys. Rev. {\bf B70} (2004) 184408; 
cond-mat/0409310.

\bibitem{BGOF04}
M. T. Batchelor, X. W. Guan, N. Oelkers, A. Foerster: 
JSTAT (2004) P10017; 
cond-mat/0409311.

\bibitem{PS81}
        J. H. H. Perk and C. L. Schultz: 
        Phys. Lett. {\bf 84A} (1981) 407-410.

\bibitem{Sc83}
 C. L. Schultz: Physica {\bf A122} (1983) 71-88.
 
\bibitem{algebra}
$U_{q}(\widehat{sl(r+1)})$ is an algebra for 
a symmetry of the $R$-matrix of the model. 
See for example, M. Jimbo: Commun. Math. Phys. {\bf 102} 537--547.
 We assume that $q=e^{\eta}$ is not a root of unity.

\bibitem{U70}
 G.V. Uimin: JETP Lett. {\bf 12} (1970) 225.
 
\bibitem{S75}
B. Sutherland: Phys. Rev. {\bf B12} (1975) 3795-3805. 
        
\bibitem{TS72} 
M. Takahashi and M. Suzuki:  
 Prog. Theor. Phys. {\bf 48} (1972) 2187.
         
\bibitem{KSS98} 
A. Kuniba, K. Sakai and J. Suzuki:
Nucl. Phys. {\bf B525} (1998) 597-626; 
math.QA/9803056.  

\bibitem{SI87} M. Suzuki and M. Inoue: 
       Prog. Theor. Phys. {\bf 78} (1987) 787--799.
         
\bibitem{K87}
        T. Koma:
 Prog. Theor. Phys. {\bf 78} (1987) 1213--1218.

\bibitem{SAW90} 
        J. Suzuki, Y. Akutsu and M. Wadati: 
        J. Phys. Soc. Jpn. {\bf 59} (1990) 2667--2680.

\bibitem{KWZ97}
A. Kl\"umper, T. Wehner and J. Zittartz:  
J. Phys. A: Math. Gen. {\bf 30} (1997) 1897-1912. 

\bibitem{JKS98} 
G. J\"uttner, A. Kl\"umper and J. Suzuki: 
   Nucl. Phys. {\bf B512} (1998) 581-600; 
   hep-th/9707074.

\bibitem{FK99} 
A. Fujii and A. Kl\"umper:  
       Nucl. Phys. {\bf B546} (1999) 751-764; 
cond-mat/9811234.
 
\bibitem{r-index} 
The lower index $i,j$ of $R_{ij}(v)$ is used in 
conventional manner. For example, suppose $E^{a,b}_{k}$ 
is defined on ${\mathbb C}_{r+1}^{\otimes (L+1)}$:  
$E^{a,b}_{k}=I^{\otimes k}\otimes E^{a,b}\otimes I^{\otimes (L-k)}$ 
where $I$ is $r+1$ by $r+1$ identity matrix; 
$k=0,1,\dots, L$. 
Then 
$R_{ij}(v)$ is defined as 
$
R_{ij}(v)=\sum_{a_{1},a_{2},b_{1},b_{2}} 
R^{a_{1},b_{1}}_{a_{2},b_{2}}(v) 
E^{a_{1},a_{2}}_{i} E^{b_{1},b_{2}}_{j}
$.

\bibitem{BVV82}
O. Babelon, H. J. de Vega and C-M. Viallet:  
Nucl. Phys. {\bf B200} (1982) 266-280. 

\bibitem{aba} 
This is a kind of \symbol{"60}dress universality' in the analytic Bethe ansatz: 
A. Kuniba and J. Suzuki: Commun. Math. Phys. {\bf 173}(1995) 225--264;  
hep-th/9406180; N. Yu. Reshetikhin: Sov. Phys. JETP {\bf 57} (1983) 691--696.
       
\bibitem{alge-ba}   
For $r=1$ case, the algebraic Bethe ansatz for the QTM is 
similar to the one for the row-to-row transfer matrix. 
See for example, 
F. G\"ohmann, A. Kl\"umper and A. Seel:  
J. Phys. A: Math. Gen. {\bf 37} 7625-7651; hep-th/0405089.  
         

\bibitem{BR90}
        V. V. Bazhanov and N. Reshetikhin: 
     J. Phys. A Math. Gen. {\bf 23} (1990) 1477--1492.

\bibitem{KRS81}
P. P. Kulish, N. Yu. Reshetikhin and E. K. Sklyanin: 
Lett. Math. Phys. {\bf 5} (1981) 393--403.

\bibitem{q-sys}
This quantity is related to a solution of a 
system of functional relations, called 
$Q$-system: 
A. N. Kirillov:  
J. Sov. Math. {\bf 47} (1989) 2450--2459; 
A. N. Kirillov and N. Yu. Reshetikhin,  
        {\it J. Sov. Math.} {\bf 52} (1990) 3156--3164.

\bibitem{DV95}
C. Destri and H. J. de Vega:  Nucl. Phys. {\bf B438} 
[FS] (1995) 413-454; hep-th/9407117. 

\bibitem{RST02}
O. Rojas, S. M. de Souza and M. T. Thomaz: 
J. Math. Phys. {\bf 43} (2002) 1390-1407; 
hep-ph/0012368.


\bibitem{BEU00}
A. B\"uhler, N. Elstner and G. S. Uhrig:
 Eur. Phys. J {\bf B16} 
(2000) 475-486.
\end{thebibliography}
\end{document}